\documentstyle{aipproc}

\begin{document}
\title{Value of the Cosmological Constant: Theory versus 
Experiment}

\author{Moshe Carmeli and Tanya Kuzmenko}
\address{Department of Physics, Ben Gurion University, Beer Sheva 84105, 
Israel}
\maketitle
\begin{abstract}
The numerical value of the cosmological constant is calculated using a 
recently suggested cosmological model and found to be $\Lambda=2.036\times 
10^{-35}$s$^{-2}$. This 
value of $\Lambda$ is in excellent agreement with the measurements recently
obtained by the {\it High-Z Supernova Team} and the {\it Supernova Cosmology 
Project}.
\end{abstract}
The problem of the cosmological constant and the vacuum energy associated with
it is of high interest these days. There are many questions related to it at
the quantum level, all of which are related to quantum gravity. Why there 
exists the critical mass density and why the cosmological constant has this
value? Trying to answer these questions and others were recently the subject 
of many publications \cite{1,2,4,6,7,9,10,12,13,15,16,17,18}.

In this paper it is shown that the recently suggested cosmological model \cite{19} 
predicts the value $\Lambda=2.036\times 10^{-35}$s$^{-2}$ for the 
cosmological constant. This value of $\Lambda$ is in excellent agreement with
the measurements recently obtained by the {\it High-Z Supernova Team} and the
{\it Supernova Cosmological Project} \cite{20,21,22,23,24,25,26}.

The Einstein gravitational field equations  
with the added cosmological term are \cite{27}:
\begin{equation}
R_{\mu\nu}-\frac{1}{2}g_{\mu\nu}R+\Lambda g_{\mu\nu}=\kappa T_{\mu\nu},
\end{equation}
where $\Lambda$ is the cosmological constant, the value of which is supposed to
be determined by experiment. In Eq. (1) $R_{\mu\nu}$ and $R$ are the Ricci 
tensor and scalar, respectively, $\kappa=8\pi G$, where $G$ is Newton's constant
and the speed of light is taken as unity.

Recently the two groups (the {\it Supernovae Cosmology Project} and the {\it 
High-Z Supernova Team}) concluded that the expansion of the universe is 
accelerating \cite{20,21,22,23,24,25,26}. 
Both teams obtained
\begin{equation}
\Omega_M\approx 0.3,\hspace{5mm} \Omega_\Lambda\approx 0.7,
\end{equation}
and ruled out the traditional ($\Omega_M$, $\Omega_\Lambda$)=(1, 0)
universe. Their value of the density parameter $\Omega_\Lambda$ corresponds to
a cosmological constant that is small but, nevertheless, nonzero and positive,
\begin{equation}
\Lambda\approx 10^{-52}\mbox{\rm m}^{-2}\approx 10^{-35}\mbox{\rm s}^{-2}.
\end{equation}

In Ref. 14 a four-dimensional cosmological model was presented. The model
predicts that the universe accelerates and hence it is equivalent to having
a positive value for cosmological constant in it.  
In the framework of this model the 
zero-zero component of Einstein's equations is written as
\begin{equation}
R_0^0-\frac{1}{2}\delta_0^0R=\kappa\rho_{eff}=\kappa\left(\rho-\rho_c^{BC}
\right)
\end{equation}
where $\rho_c^{BC}=3/\kappa\tau^2$ is the critical mass density
and $\tau$ is Hubble's time in the zero-gravity limit.

Comparing Eq. (4) with the zero-zero component of Eq. (1), one obtains the
expression for the cosmological constant,
\begin{equation}
\Lambda=\kappa\rho_c^{BC}=3/\tau^2.
\end{equation}

To find out the numerical value of $\tau$ we use the relationship between
$h=\tau^{-1}$ and $H_0$ given in Ref. 14 [Eq. (5.23)]:
\begin{equation}
H_0=h\left[1-\left(1-\Omega_M^{BC}\right)z^2/6\right],
\end{equation}
where $z$ is the redshift and $\Omega_M^{BC}=\rho_M/\rho_c^{BC}$ where 
$\rho_c^{BC}=3h^2/8\pi G$ \cite{19}. (Notice that $\rho_c^{BC}$ is different from 
the standard $\rho_c$ defined with $H_0$.) The redshift parameter $z$ 
determines the distance at which $H_0$ is measured. We choose $z=1$ (Fig. 11
in Ref. 14) and take for 
\begin{equation}
\Omega_M^{BC}=0.245
\end{equation}
(roughly corresponds to 0.3 in the standard theory), Eq. (6) then gives
\begin{equation}
H_0=0.874h.
\end{equation}
At the value $z=1$ the corresponding Hubble constant $H_0$ according to the 
latest results from HST can be taken \cite{28} as $H_0=72$km/s-Mpc, thus 
$h=(72/0.874)$km/s-Mpc or
\begin{equation}
h=82.380\mbox{\rm km/s-Mpc},
\end{equation}
and 
\begin{equation}
\tau=12.16\times 10^9 \mbox{\rm years}.
\end{equation}

What is left is to find the value of $\Omega_\Lambda^{BC}$. We have 
$\Omega_\Lambda^{BC}=\rho_c^{ST}/\rho_c^{BC}$, where $\rho_c^{ST}=3H_0^2/8\pi 
G$ and $\rho_c^{BC}=3h^2/8\pi G$. Thus $\Omega_\Lambda^{BC}=(H_0/h)^2=0.874^2$,
or
\begin{equation}
\Omega_\Lambda^{BC}=0.764.
\end{equation}
As is seen from Eqs. (7) and (11) one has 
\begin{equation}
\Omega_M^{BC}+\Omega_\Lambda^{BC}=1.009\approx 1,
\end{equation}
which means the universe is flat.

As a final result we calculate the cosmological constant according to Eq. (5).
One obtains
\begin{equation}
\Lambda=3/\tau^2=2.036\times 10^{-35}s^{-2}.
\end{equation}

Our results confirm those of the supernovae experiments and indicate on
existance of the dark energy as has recently received confirmation from the
Boomerang cosmic microwave background experiment \cite{29,30}, which showed that 
the universe is flat.

\end{document}